\documentclass[11pt,twoside]{article}  
\usepackage{asp2006}
\usepackage{adassconf}

\newcommand{\tgcat}{\textbf{\textit{TGC}at} } 
\newcommand{\tgcatnsp}{\textbf{\textit{TGC}at}} 

\begin{document}   

%
%

\paperID{P45}

%

\title{\tgcatnsp, The Chandra Transmission Grating Catalog and Archive: Systems, Design and Accessibility}

%
%
%
%
%

\markboth{MITSCHANG, HUENEMOERDER, NICHOLS}{\tgcatnsp}

%
%
%
%

\author{Arik W. Mitschang\altaffilmark{1},
        David P. Huenemoerder\altaffilmark{2},
        Joy S. Nichols\altaffilmark{1}} 
        
\altaffiltext{1}{Smithsonian Astrophysical Observatory, Cambridge, MA, USA}
\altaffiltext{2}{MIT Kavli Institute for Space Research, Cambridge, MA}

%

\contact{Arik Mitschang}
\email{amitschang@cfa.harvard.edu}

%
%
%

\paindex{Mitschang, A. W.}
\aindex{Huenemoerder, D. P.}
\aindex{Nichols, J. S.}


%

\keywords{}


\begin{abstract}          
 The recently released Chandra Transmission Grating Catalog and
 Archive, \tgcatnsp, presents a fully dynamic on-line catalog allowing
 users to browse and categorize Chandra gratings observations quickly
 and easily, generate custom plots of resulting response corrected
 spectra on-line without the need for special software and to download
 analysis ready products from multiple observations in one convenient
 operation. \tgcat has been registered as a VO resource with the NVO
 providing direct access to the catalogs interface. The catalog is
 supported by a back-end designed to automatically fetch newly public
 data, process , archive and catalog them, At the same time utilizing
 an advanced queue system integrated into the archive's MySQL database
 allowing large processing projects to take advantage of an unlimited
 number of CPUs across a network for rapid completion. A unique
 feature of the catalog is that all of the high level functions used
 to retrieve inputs from the Chandra archive and to generate the final
 data products are available to the user in an ISIS written library
 with detailed documentation. Here we present a structural overview of
 the Systems, Design, and Accessibility features of the catalog and
 archive.
\end{abstract}

%
%

\section{Introduction}
\tgcat aims to be the definitive end-user source for all Chandra HETGS
and LETGS observations. In order to achieve this goal the catalog must
both have the absolute best processed and calibrated data, as well as
have an interface which makes it easy for users to find, review and
download their observations of choice. The science requirements for
accurately processing gratings data are discussed elsewhere (
Huenemoerder et.\ al.\ 2010 ).
This writing will focus on the automated system that collects and
processes new public data, the interfaces used for administrative
review, and the interfaces and systems provided for user access.

\section{Database, Archive,  and Subsystems}

The \tgcat processing system is comprised of three major components:
\begin{itemize}
\item MySQL database storing meta-data
\item File archive
\item Processing Software
\end{itemize}
These components are described in the context of \tgcat processing in
this section.

\paragraph{Tables:}
The tables relevant to processing include the extractions, source,
spectral properties, files, and queue tables. The extractions table
has one entry per processed extraction, where extraction is taken to
mean a single source in a unique \textit{Chandra} observation ID
(ObsID). Any one ObsID can have many sources and any one source can be
in many ObsIDs, the extractions table will store one entry for each
combination thereof. In order to consolidate all extractions of a
single source, there is a source table indexed on
\htmladdnormallinkfoot{SIMBAD}{http://simbad.u-strasbg.fr/simbad}
identifier, a \tgcat identifier, and coordinates, that associates
entries in the extractions table. The files table tracks processing
output products and summary images allowing the web pages to easily
display information on files available for download. The spectral
properties table stores and indexes spectral properties in several
different wavebands, these data are also available in a fits table for
download.

\paragraph{Data Flow:}

\begin{figure}[t]
\epsscale{1}
\plotone{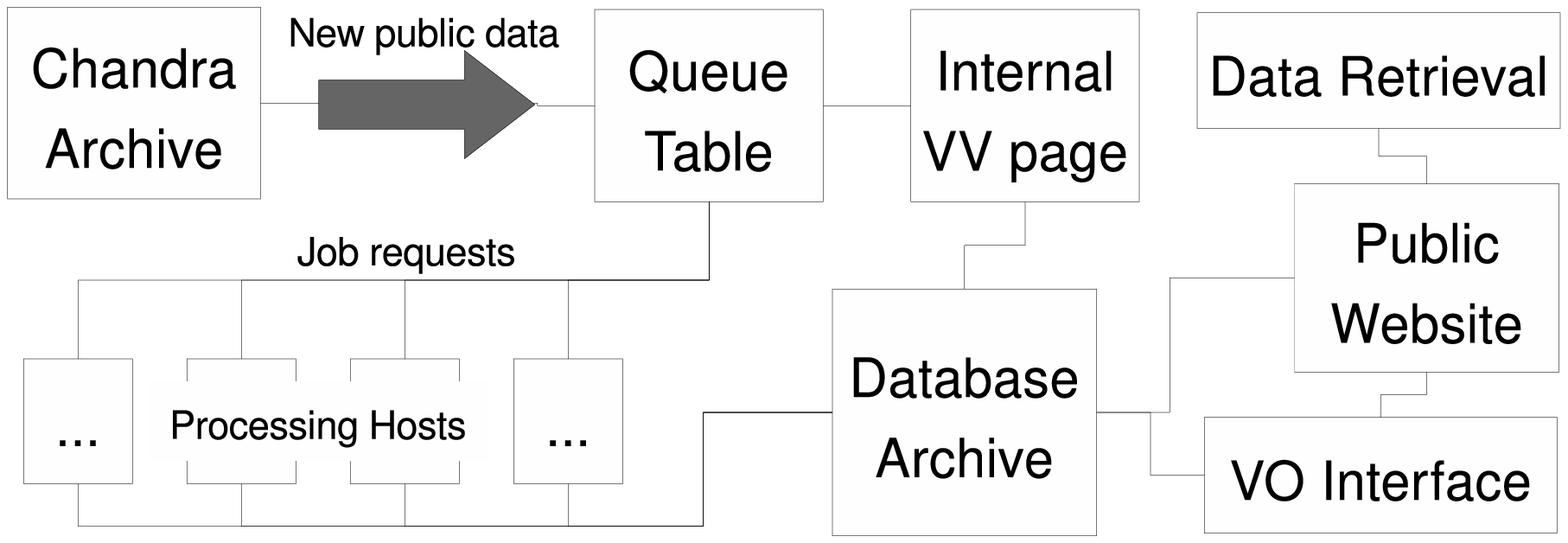}
\caption{\tgcat data flow} 
\label{systems_diag}
\end{figure}

The flow of data through the system is rather simple, as illustrated
in Figure \ref{systems_diag}.
%
%
A bash script run via \textit{cron} at regular intervals downloads a
list of gratings observations from the public chandra
\htmladdnormallinkfoot{archive}{http://cda.harvard.edu/chaser/}
and compares with the list of ObsIDs that have been submitted to \tgcatnsp.
Any not in \tgcat will be added to the queue table to be processed in
line. Daemon processes, written in python, run on any number of
network connected hosts continuously requesting entries from the queue
in a \textit{FIFO} manner, the first process to ask entries from the
queue table will retrieve the first entry ( by time of creation
). They then parse the queue entry which can contain a number of
custom processing parameters, setup work spaces and logging, fork off
the processing which is implemented in ISIS ( Houck \& Denicola 2000 )
%
%
interactive library functions, which are available for
\htmladdnormallinkfoot{download}{http://space.mit.edu/cxc/analysis/tgcat}
and custom use, and finally ingest data into the database and archive.
During the ingest step, several checks are done to evaluate whether or
not the resultant processed data are a complete set and worth being
manually reviewed by \tgcat scientists. If so, meta data are added to
the database's main extractions table which returns a unique identifier
that is then used to tag the file based products. At this time linking
is done between the source table and extractions table.  If not, the
queue entry is marked as an error and notification sent.  \tgcat
operators have the choice to investigate the existing processing
workspace or simply re-evaluate parameters and re-queue the extraction
as new.

\paragraph{Validation and Verification:}
Each extraction available for browsing has been reviewed by a member
of the \tgcat science team in order to confirm zeroth order placement,
proper masking, etc. This is done via an internal website nearly
mimicking the public interface for reviewing extractions but with the
addition of forms for queueing processing and for marking the
extraction as ``good'', ``bad'', or ``warning'' and optional comment
fields which are available for review by the end user. ``bad''
extractions are never shown on the public site and administrators have
the option of rejecting any other extraction for whatever
reason. Because each extraction is tagged with the date of processing,
the version of \tgcat used for processing, and a so called group ID
which is unique per group of extractions intended to be of the same
object for a single ObsID, keeping track of accepted sources and
avoiding duplicates is made easy.

\section{Access}
\tgcat has several interfaces for data access. Currently three are
operational including the web browser interface, and two Virtual
Observatory access protocol interfaces. Each of these access
interfaces are implemented as a ``plug-in'' written in php. Plug-in
functions to create output are called by a catalog independent
generalized query library written in php under \tgcat development
dubbed ``queryLib''. This modular approach makes it trivial to add
access interfaces to \tgcatnsp.

\paragraph{queryLib:}
queryLib stores state and type information of an individual query in a
database table. Each time a query is performed, fields in this table
such as a \textit{where} clause, sort fields, indexed IDs and columns
are populated with information specific to that query. As a user
interacts with the particular interface, information in this table
entry are updated to reflect the new state of the query through object
method calls. queryLib initialized a query by creating a new entry in
the table with two unique IDs: one is the primary key identifier
assigned at creation and can be used to reference the query, the other
is created on the fly using a combination of the primary key and a
random hash string that is very unlikely to be guessed.  This is done
so that a malicious user cannot change the state of anyone's query by
simply guessing an ID. The queryLib speeds up the process of redrawing
a table of results by storing the indexed primary keys of all entries
selected rather than having to rerun a potentially complicated where
clause.
%
%
\paragraph{Web Catalog Interface:}
The web \htmladdnormallinkfoot{interface}{http://tgcat.mit.edu} to
\tgcat allows users browse extraction and object entries in a way that
is most suitable to the catalog data structure. Each extraction has an
associated preview page where a set of standard plots produced at the
time of processing is available for viewing. A table of spectral
properties
%
%
providing quick general categorization, and a table of each individual
file available for the extraction along with a link to download that
file are provided. Perhaps one of the most important features of
the web interface is the interactive plotting of individual or
combined spectra directly on the site. This is implemented server side
by taking the POST request for plotting parameters along with a unique
file name, creating a small script of the commands for loading data and
plotting, and piping this script into an ISIS process. Malicious use
is prevented by checking all parameter inputs for appropriate values,
running the ISIS process as an unprivileged user, and checking the
temp file name for validity before piping. Since the file name is known
at the time of the request, the page simply needs to reload to show
the new plot. The commands, any error output, and an ASCII dump of
the plot data are available for download as well.
\paragraph{Virtual Observatory:}
\tgcat is a registered VO service providing both the
\htmladdnormallinkfoot{Simple Cone Search}{http://tgcat.mit.edu/tgCli.php?OUTPUT=V} 
and 
\htmladdnormallinkfoot{Simple Image Access}{http://tgcat.mit.edu/tgSia.php} 
Protocols. These are both implemented as an XML output plug-ins taking
input from a php script that provides appropriate GET parameters, then
creates a query object after parsing the input data, finally running
the query in exactly the same way as for the web and ASCII
interfaces. Error handling is done by the calling script and meta data
is provided by running a query with parameters known to have no
catalog entries ( indexed ID=0 ). VO queries, and any other type, can
be tracked in the query table using the type column. In this way we
can track statistics on requests coming from services such as
\htmladdnormallinkfoot{datascope}{http://heasarc.gsfc.nasa.gov/cgi-bin/vo/datascope/init.pl}.
\paragraph{Data Package Downloads:}
Generally, users will want to download more than one file at a time
for analysis, such as such as spectra and responses or products from
multiple extractions of the same target. To this end \tgcat runs a
packager process that parses a queue table much like the processing
queue table described above. Requested packages are added to the
package queue table, validated and read by the packager which then
fetches data to a temp space placing them in a directory hierarchy
tagged with ObsID and \tgcat ID. The entire hierarchy is then
tarred, compressed and provided to the user along with file checksums
via HTTP.

\paragraph{Acknowledgements.} 
We would like to thank Dan Dewey for extensive input on data
organization and interface layout, and Mike Nowak for ISIS plotting
routines and advice on the interactive plotting interface. This work
is supported by the Chandra X-ray Center (CXC) NASA contract
NAS8-03060. DPH was supported by NASA through the Smithsonian
Astrophysical Observatory (SAO) contract SV3-73016 for the Chandra
X-Ray Center and Science Instruments.


\end{document}